\documentclass[preprint]{aastex}
\usepackage{emulateapj5}

\newcommand{\etal}{\mbox{et al.}}
\newcommand{\ksxrb}{\mbox{KS 1731$-$260}}
\newcommand{\aqlxone}{\mbox{Aql X-1}}
\newcommand{\mxbgc}{\mbox{MXB 1743$-$29}}

\newcommand{\lmxbs}[2]{\mbox{4U #1$-$#2}}

\newcommand{\rxte}{{\it RXTE}}

\newcommand{\nco}{burst oscillations}

\begin{document}

\shortauthors{Muno et al.}
\shorttitle{Burst Oscillations and Radius Expansion}

%\submitted{Submitted to ApJ Letters}
\title{Millisecond Oscillations and Photospheric Radius Expansion
in Thermonuclear X-Ray Bursts}
\author{Michael P. Muno, Deepto Chakrabarty, Duncan K. Galloway, and
Pavlin Savov}
\affil{Department of Physics and Center for Space Research, 
       Massachusetts Institute of Technology, Cambridge, MA 02139}
\email{muno,deepto,duncan,pavlin@space.mit.edu}

\begin{abstract}
We use archival data from the {\em Rossi X-Ray Timing Explorer} to
examine 125 type~I X-ray bursts from the 9 weakly magnetic accreting
neutron stars where millisecond oscillations have been detected during
some bursts.  We find that oscillations from the 6 ``fast'' ($\simeq
600$ Hz) sources are almost always observed during radius expansion bursts,
whereas oscillations from the 3 ``slow'' ($\simeq 300$ Hz) sources are
about equally likely to be found in bursts both with and without 
photospheric radius expansion.
This strongly suggests that the distinction between these two source groups
cannot be an observational selection effect, but must instead arise
from some physical mechanism.    
\end{abstract}

\section{Introduction}

Nearly coherent 270--620 Hz oscillations have been observed during
type~I X-ray bursts\citep{lvt93} from neutron stars in nine low mass X-ray binaries
(LMXBs; see van der Klis 2000 for a recent review).  
Oscillations may occur because a hot spot forms on the rotating
neutron star surface during the burst \citep{str96}. Their frequency
evolution has been interpreted in terms of a burning 
layer that expands by $\approx$50~m and slows at the start of the
burst, only to spin up again over several seconds as the layer
contracts \citep{str97, cb00}.  

The pairs of kilohertz quasi-periodic oscillations (kHz QPOs; see van
der Klis 2000 for a review) observed in the persistent emission of
most burst oscillation sources naturally divide the sources into two
categories.  The frequencies of these twin kHz QPOs vary by up to a
factor of 2, while their separation $\Delta\nu_{\rm kHz}$ varies by at
most $\sim 40$\% and suggests the relation $\nu_{\rm burst} \approx
n\,\Delta\nu_{\rm kHz}$, with $n=1$ for the three sources with a
``slow'' $\nu_{\rm burst} \simeq 300$~Hz and $n=2$ for the six sources
with a ``fast'' $\nu_{\rm burst} \simeq 600$~Hz \citep{str96, vdk00}.
This relationship between $\Delta\nu_{\rm kHz}$ and $\nu_{\rm burst}$
has been interpreted in terms of a beat frequency model for the kHz
QPOs relating the inner accretion disk motion with the neutron star
spin frequency, $\nu_{\rm spin}$ \citep{str96, mlp98}.  This model
presumes that $\Delta\nu_{\rm kHz}\approx\nu_{\rm spin}$ and suggests
that sources with slow \nco\ ($n=1$) have a single visible hot spot on
their surface and that sources with fast \nco\ ($n=2$) have a pair of
hot spots \citep{str98a}.  This further suggests that all nine sources have
$\nu_{\rm spin}\simeq 300$~Hz, and several authors have proposed
mechanisms for a natural spin equilibrium near this frequency (White
\& Zhang 1997; Bildsten 1998).  An alternative explanation for the
kHz QPOs in terms of relativistic accretion-disk precession has also
been proposed (Stella \& Vietri 1998; Psaltis \& Norman 2000), but it
does not explicitly address the relation between $\Delta\nu_{\rm kHz}$
and $\nu_{\rm burst}$. 

Millisecond oscillations do not occur in every burst from these nine
LMXBs.  \citet{mun00} found that oscillations from \ksxrb\ are 
usually observed only in bursts which exhibit photospheric radius
expansion, during which  the apparent radius of the emission region
increases by $\ga 20$~km because the flux at the surface of the neutron
star exceeds the Eddington limit.   
On the other hand, \citet{fra00} and \citet{vst00} found that bursts
from \lmxbs{1728}{34} {\it without} radius expansion exhibited
oscillations more often than radius expansion bursts.  One
common property of both sources is that millisecond oscillations are
only observed from bursts which occur on the so-called ``banana''
branch of their X-ray color-color diagrams, which corresponds to
relatively high accretion rates (see van der Klis 1995 for a review
of LMXB phenomenology).  In this {\em Letter}, we explore 
these dependencies of the burst oscillation phenomenon on burst 
properties by making a comparison between X-ray bursts with and without
coherent oscillations in all 9 sources using archival data from the
{\it Rossi X-ray Timing Explorer} (\rxte). 

\section{Observations and Data Analysis}

We have examined all observations of 8\footnote{Since
the localization of bursts associated with \mxbgc\ is uncertain due to
its proximity to the bursting pulsar GRO J1744$-$28, we use the
results from Strohmayer et al. (1997a). For the same reason, the persistent 
emission cannot be isolated for this source.} of the 9 burst oscillation
sources in the {\em RXTE} public archive %\footnote{Available at the
%High Energy Astrophysics Science Archive Research Center (HEASARC) at
%NASA Goddard Space Flight Center, http://heasarc.gsfc.nasa.gov.} 
as of 2001 March 22.  We searched all of the data taken with the {\em
RXTE} Proportional Counter Array (PCA) for type~I bursts and found a
total of 125 bursts with data that allowed both timing and spectral
analyses. These bursts are summarized by source in Table~1. 
For the five sources
for which detailed burst studies have been published, our results are
consistent with the published work (\lmxbs{1608}{52}: Chakrabarty et
al. 2001; \lmxbs{1658}{298}: Wijnands, Strohmayer, \& Franco 2001; 
\aqlxone: Fox et al. 2001; \ksxrb: Muno et al. 2000;
\lmxbs{1728}{34}: Franco 2001, van Straaten et al. 2001; and
\lmxbs{1916}{05}: Galloway et al. 2001).

In order to characterize the persistent emission, we
calculated average soft (3.7--5.1 keV/2.3--3.7 keV) and hard (8.7--18
keV/5.1--8.7 keV) X-ray colors for 256~s intervals
using background-subtracted light curves.  
Location on an X-ray color-color diagram is a good tracer of
the accretion rate $\dot M$ (see van der Klis 1995 and references
therein).  \ksxrb\
\citep{mun00} and \lmxbs{1728}{34} \citep{fra00,vst00} exhibit bursts
over the widest range of $\dot M$, while cursory checks of the
color-color diagrams for the other sources suggest that bursts were
only observed at high accretion rates because few or no observations
were made during low $\dot M$ intervals.

To search for oscillations, we created twice-oversampled
power spectra for 2~s intervals of data every 0.25~s
for the duration of each burst.  If oscillations are detected within
$\pm 3$~Hz of the expected frequency and within 10 seconds after the
start of a burst with a probability $< 2\times10^{-4}$
that the signal is due to Poisson noise in a single trial, 
we consider the oscillation to be a detection.  We also
used this technique to search all of the bursts for oscillations at
the first three harmonics of the slow oscillations, and at 
0.5, 1.5, and 2.0 times the frequency of the fast oscillations. We 
found no evidence for oscillations at these frequencies in 
any of the bursts. 
Using power spectra of 1 s intervals, we can place upper limits on the 
fractional RMS amplitude of
oscillations at these frequencies of 5--15\% during a burst.
%
% for +-3 Hz, 10 s, 99\% threshold is 17.7 
%
These are not very stringent upper limits, so application of more
sensitive search techniques such as those described in \citet{mil99},
\cite{mun00}, and Strohmayer (2001) will be useful for more detailed studies.  

We produced energy spectra for each 0.25 s interval from each burst
using available combinations of data modes which provide at least 32 energy
channels.  We subtracted spectra from 15 s of emission from 
before the burst 
to account for background, and fit each spectrum between 2.5--20 keV 
with a model consisting of a
blackbody multiplied by a constant interstellar absorption (determined
from the mean value from fits using variable absorption). 
The model provides an apparent temperature ($T_{\rm app}$) and a
normalization equal to the square of the apparent radius ($R_{\rm
app}$) of the burst emission surface, and allows us to estimate the
bolometric flux as a function of time.  The peak flux of bursts from a given
source can vary by a factor of 5--10.
In many bursts, photospheric radius expansion is evident at the start
of the burst, during which $R_{\rm app}$ increases and $T_{\rm app}$
decreases such that the bolometric flux remains constant, presumably
at the Eddington limit (see Lewin et al. 1995). Our definition of 
radius expansion includes bursts which exhibit a second increase in radius
immediately after the minimum which follows the expansion phase. We find 
several such bursts from \lmxbs{1728}{34} \citep[see also][]{vst00}, 
and one such burst from both \lmxbs{1916}{053} and \lmxbs{1608}{52}. 

\section{Results}

A summary of our results for the nine burst oscillation sources is
given in Table~1. 
There is a tight connection between
the presence of oscillations and of radius expansion
in fast ($\simeq 600$~Hz) sources from every perspective.
Fast oscillations occur predominantly during bursts with radius expansion,
and almost all bursts with radius expansion exhibit oscillations. 
At the same time most of the bursts without fast oscillations
also lack radius expansion. In
slow ($\simeq 300$~Hz) sources, there is no preference for whether 
bursts with or without radius expansion exhibit oscillations.
This suggests that the frequencies of
burst oscillations and the properties of bursts are connected.  

\begin{figure}
\plottwo{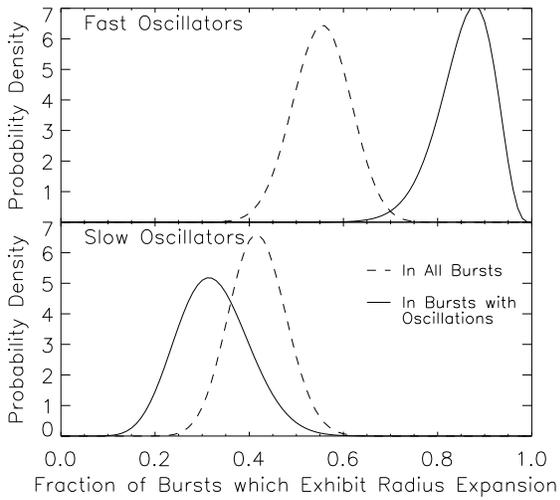}{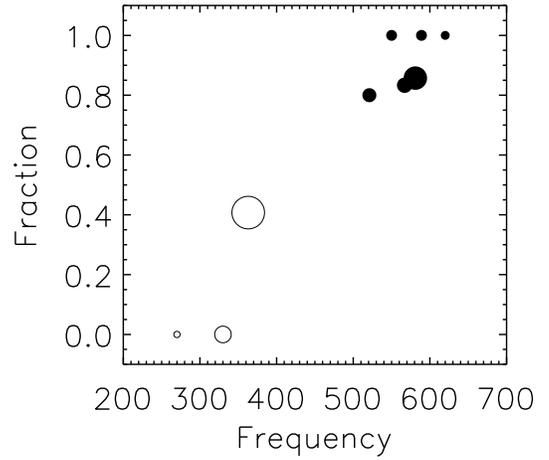}
\caption{The probability that radius expansion bursts occur
in a given fraction of all bursts ({\it dashed line}) and 
of bursts in which oscillations are observed ({\it solid line}),
assuming that the number of radius expansion bursts out of a sample 
population is distributed according to a binomial distribution. 
\label{dist}}  
\caption{The fraction of bursts with oscillations that are also
radius expansion bursts, plotted as a function of source oscillation
frequency. Open circles denote slow ($n=1$) oscillation sources, while
solid circles denote fast ($n=2$) oscillation sources. The size of the
circle indicates the number of bursts with oscillations observed for
each source.
\label{whendo}}  
\end{figure}

Although the sample of bursts from an individual source is in some cases
quite small, the correlations for fast and slow sources as groups are quite 
significant. In order to quantify the significance of our results, we 
hypothesize that radius expansion occurs in bursts with a fixed 
probability, $f$. Given a sample of $m$ bursts, 
the probability of observing a
number $n$ of bursts with radius expansion is 
$$P(n|f, m) = f^n (1-f)^{m-n} {{m!} \over {n! (m-n)!}}.$$ 
We can then compute the probability density for $f$
given $n$ radius expansion bursts in a sample of $m$ bursts, 
$p(f| n, m)$. We have plotted these probability densities in 
Figure~\ref{dist} for fast and slow sources, considering as our sample
population either all observed bursts (dashed line) and only those bursts which 
exhibit oscillations (solid line). Although fast and slow sources exhibit
radius expansion in about equal fractions of bursts in general, 
fast sources exhibit radius expansion during bursts with 
oscillations far more often (88$^{+4}_{-8}$\% of the time) than 
slow sources (31$^{+10}_{-6}$\%).
The trends for individual sources are interesting in that they 
follow the behavior expected from considering the sources as groups
(Figure~\ref{whendo}). On the other hand,
slow sources are more likely to show radius expansion during bursts 
{\it without} oscillations (53$^{+9}_{-8}$\%) than fast sources 
(20$^{+10}_{-5}$\%).

We can rule out some observational selection effects as causes for these 
correlations. For example, we are not
systematically missing oscillations from weak bursts, as might occur
if all oscillations had the same fractional amplitude. We observe
oscillations in some of the weakest bursts from \ksxrb,
\lmxbs{1636}{53}, \lmxbs{1916}{053}, and \lmxbs{1728}{34}, while no
oscillations are detected during some of the stronger bursts from
these sources. Therefore, our correlations are based upon genuine
variations in the strengths of the oscillations.
 
However, since bursts were observed from only two sources at low 
$\dot M$ (\ksxrb\ and \lmxbs{1728}{34}), there is a remote 
chance that 
these correlations are an artifact of the higher $\dot M$
at which the remaining sources were observed. For instance, the bursts 
with the longest time scales take place at low fluxes in \lmxbs{1608}{52} 
\citep{mur80} and at low inferred $\dot M$ in \lmxbs{1636}{536}
\citep{vdk90}. It is reasonable to believe that these long bursts do not
exhibit radius expansion (as in \ksxrb; see for example
Muno et al. 2000), so there
would be no strict relationship between radius expansion and
fast oscillations if these bursts exhibit oscillations. 
However, we would consider this a surprise given 
the absence of oscillations at low accretion rates in \ksxrb\ 
and \lmxbs{1728}{34}. On the 
other hand, the time scales of bursts from the slow oscillators 
\lmxbs{1916}{053} (Swank, Taam, \& White 1984) and \lmxbs{1702}{429} \citep{mak82}
 have not been observed to vary systematically with the persistent flux, 
so we cannot predict how observing bursts
at low $\dot M$ in slow sources would affect our correlations.

\section{Discussion}

We have found that oscillations from the 6 fast burst oscillation 
sources are tightly 
connected to photospheric radius expansion, whereas oscillations from the 3 slow 
sources are about equally likely to be found in bursts both with 
and without radius expansion.
What drives this correlation remains to be determined: is it the burst
properties themselves, the oscillation frequencies, or some unseen
third parameter? 

According to the  beat frequency model of kHz QPOs, the 
fact that $\nu_{\rm burst}\simeq \Delta\nu_{\rm kHz}$ 
for the ``slow'' sources whereas $\nu_{\rm burst}\simeq
2\Delta\nu_{\rm kHz}$ for the ``fast'' ones can be accounted for if
one or two antipodal hot spots on the surface of the rotating 
neutron star are visible to the observer (Miller et al.\ 1998). 
One possibility is that the distinction between fast and slow oscillations
is due to a difference in the orientation of the hot spots and the 
observer with respect to the rotation axis of the star. This seems unlikely.
Radius expansion is observed with similar likelihood from both
fast and slow sources, and therefore is unlikely to depend on our
viewing angle. We would not 
expect oscillations to be associated with radius expansion bursts 
only in the fast sources if viewing angle effects determine whether one or 
two spots are observed. 

It also does not appear that the strengths of the bursts
determine the oscillation frequencies by igniting either one or two hot spots.
If this were the case, one would expect to detect slow oscillations during weak
bursts without radius expansion from the fast sources, and fast
oscillations during strong bursts from the slow sources. Out of the
125 bursts we observed from sources of \nco, we find no evidence for
harmonic or half-frequency signals with powers comparable to the
signals at the frequencies in Table~\ref{sum}.

If the distinction between
slow and fast oscillators is equivalent to a division between slow and
fast rotators, $\nu_{\rm spin}$ (or some related
quantity, e.g., the effective surface gravity) could determine which
bursts show oscillations.  However, that option is not free of
complications, as the transition between the burst properties
for sources that exhibit fast and slow oscillations must be very sharp, 
since the two populations are not at all well separated in frequency 
(see Table~\ref{sum}).  When comparing the observed distribution of
$\nu_{\rm burst}$ to a uniform distribution of frequencies between
250--650~Hz, a Kolmogorov-Smirnov test (e.g., Eadie et al.\ 1971) can
exclude a uniform distribution at only the 1.3$\sigma$ (81\%)
confidence level.  It is interesting to note that the recent report of
a possible $\approx 400$~Hz burst oscillation from the 401~Hz pulsar
SAX J1808.4$-$3658 (in 't Zand et al. 2001) would make
the $\nu_{\rm burst}$ distribution even more consistent with a uniform
distribution (excluded at only the 0.9$\sigma$ or 64\% confidence
level), so the putative transition would have to be correspondingly
sharper.

We have listed a few additional properties of these LMXBs in 
Table~\ref{sum}.  Neither the activity
level nor the long-term average accretion rate $<\dot M>$, as
determined from nearly 5 years of data from the \rxte\ All-Sky
Monitor (Levine et al. 1996), 
appears to be correlated with the frequencies of the burst oscillations.  
Fast oscillations are observed in both transient
(\lmxbs{1608}{52} and \aqlxone) and persistent (e.g., \lmxbs{1636}{53})
sources, as well as from both low $<\dot M>$ and high $<\dot M>$
sources. Orbital periods are measured for only 4 of the 9 sources, 
and range from 0.81 to 19 hours. It is apparent that these burst oscillation 
sources are an inhomogeneous group, which makes measurements of oscillations 
from other sources highly desirable.

We feel that the most likely
explanation for the observed correlations is that the burst properties 
change differently as a function of $\dot M$ in fast and slow sources.
X-ray burst theory predicts that
radius expansion should occur only at low $\dot M$ (Fujimoto, Hanawa,
\& Miyaji 1981; Ayasli \& Joss 1982).  This agrees with observations
of the slow oscillator \lmxbs{1728}{34} (Franco 2000; van Straaten et
al. 2000), but does not appear to hold true for the fast oscillators \ksxrb\
\citep{mun00}, \lmxbs{1608}{52} \citep{mur80}, and \lmxbs{1636}{53}
\citep{vdk90}.  If oscillations only appear at high $\dot M$ (as suggested by 
Franco 2001), then they would indeed be associated with radius expansion in 
the fast sources, but not the slow sources.
Furthermore, Bildsten (2000) has suggested that some mechanism acts
in these latter sources to confine the accreted material such that the 
{\em local} $\dot M$ can decrease even as the global $\dot M$ increases.  
If this is true, such confinement is somehow related to the higher frequency
of the fast burst oscillations. 

\acknowledgements{We are grateful to Dimitrios Psaltis for
providing many helpful suggestions and ideas, and to the referee for 
many comments which improved the tone and presentation of this paper. 
It is also a pleasure to 
thank Lars Bildsten, Andrew Cumming, Fred Lamb, and Cole Miller for 
useful discussions.  This work was supported by the NASA Long-Term Space 
Astrophysics program under grant NAG 5-9184, as well as by NASA contract 
NAS 5-30612.}

\begin{deluxetable}{lcccccccccc}
\tabletypesize{\scriptsize}
\tablecolumns{11}
\tablewidth{0pc}
\tablecaption{Burst Oscillations (BOs) and Photospheric Radius
Expansion (PRE)\label{sum}}
\tablehead{
 \colhead{} & \colhead{} & \colhead{} & 
   \multicolumn{4}{c}{Number (Percentage) of Bursts} & & & &  \\ \cline{4-7}
 \colhead{Source} & \colhead{$\nu_{\rm burst}$} & 
   \colhead{Total} & \colhead{BOs} & \colhead{no BOs} &
   \colhead{BOs} & \colhead{no BOs} &
   \colhead{$F_{\rm ASM}$\tablenotemark{a}} & \colhead{$D$} & 
   \colhead{$P_{\rm orb}$} &  \\
 \colhead{Name} & \colhead{(Hz)} & \colhead{Bursts} & 
   \colhead{PRE} & \colhead{PRE} & \colhead{no PRE} & \colhead{no PRE} & 
   \colhead{(c s$^{-1}$)} & \colhead{(kpc)} & \colhead{(hr)} & 
   \colhead{Ref.} }

\startdata
\cutinhead{Fast Oscillators}
\lmxbs{1608}{52} & 620 & 6 & 2 (33\%) & 0 & 0 & 4 (67\%)  & 3.36 & 4.0 & \nodata & [1,2] \\ 
\mxbgc &  589 & 3 & 3 (100\%) & 0 & 0 & 0 & \nodata & \nodata & \nodata & [3] \\ %*
\lmxbs{1636}{53} & 581 & 16 & 12(75\%) & 0 & 2 (13\%) & 2 (13\%) & 14.55 &
6.5 & 3.8 & [2,4,5]  \\
\lmxbs{1658}{298} & 567 & 15 & 5 (33\%) & 5 (33\%) & 1 (7\%) & 4 (27\%) & 0.74 & 10\tablenotemark{b} & 7.1 & [4,6] \\
\aqlxone &  550 & 10 & 3 (33\%) & 1 (10\%) & 0 & 6 (60\%) & 2.79 & 4.8 & 19 & [2,7,8] \\
\ksxrb & 521 & 13 & 4 (31\%) & 0 & 1 (4\%) & 8 (62\%) & 10.12 & 7 & \nodata & [5] \\[5pt]
Total Fast & & 60 & 26 (46$^{+7}_{-6}$\%) & 6 (10$^{+5}_{-3}$\%) & 4 (6$^{+5}_{-1}$\%) & 24 (38$^{+7}_{-5}$\%)\tablenotemark{c} & & & & \\[10pt]
\cutinhead{Slow Oscillators}
\lmxbs{1728}{34} & 363 & 49 & 11 (22\%) & 11 (22\%) & 16 (33\%) & 11 (22\%) & 6.58 & 4.3 & \nodata & [10,11,12] \\
\lmxbs{1702}{429} & 330 & 8 &  0 & 0 & 7 (87\%) &  1 (13\%) & 3.50 & 6.7 & \nodata & [4,12] \\
\lmxbs{1916}{053} & 270 & 8 & 0 & 5 (50\%) & 1 (10\%) & 2 (20\%) & 1.21 & 10 & 0.83 & [2,13,14]  \\[5pt]
Total Slow & & 65 & 11 (17$^{+6}_{-3}$\%) & 16 (25$^{+6}_{-4}$\%) & 24 (37$^{+7}_{-5}$\%) & 14 (22$^{+6}_{-4}$\%)\tablenotemark{c} & & & & \\[10pt]
\enddata
\tablenotetext{a}{Mean 1.5-12 keV count rate observed with {\em
RXTE}/ASM.}
\tablenotetext{b}{Distance was derived assuming the peak flux from the
brightest burst observed with \rxte\ represents the Eddington luminosity for
pure helium.}
\tablenotetext{c}{1-$\sigma$ uncertainties derived assuming a binomial 
distribution (see text).}
\tablerefs{[1] \citet{cha00}; [2] \citet{cs97} ; [3] \citet{str97}; [4]
current paper; [5] \citet{sm88} ; [6] \citet{cw84} ; [7] \citet{fox01}; [8] 
\citet{ci91} [9] \citet{mun00}; [10] \citet{fra00}; [11] \citet{vst00};
[12] \citet{for00}; [13] \citet{gal00}; [14] \citet{wal82}}
\end{deluxetable}

\end{document}